\documentclass[aps,twocolumn,groupedaddress,showpacs]{revtex4}

\usepackage{amsmath,amsfonts}
\usepackage{epsfig,graphicx}
\usepackage{psfrag}
\begin{document}

\preprint{HEP/123-qed}

\title[Short Title]{Generating optimal states for a homodyne Bell test}

\author{Sonja Daffer}
\email{s.daffer@imperial.ac.uk}

\author{Peter L. Knight}%
\affiliation{%
    Blackett Laboratory,
    Imperial College London,
    Prince Consort Road,
    London SW7 2BW,
    United Kingdom
    }%

\date{\today}      % It is always today \today.
\begin{abstract}
\vspace{.1in} \noindent We present a protocol that produces a
conditionally prepared state that can be used for a Bell test
based on homodyne detection.  Based on the results of Munro [PRA
1999], the state is near-optimal for Bell-inequality violations
based on quadrature-phase homodyne measurements that use
correlated photon-number states. The scheme utilizes the Gaussian
entanglement distillation protocol of Eisert \textit{et.al.}
[Annals of Phys. 2004] and uses only beam splitters and
photodetection to conditionally prepare a non-Gaussian state from
a source of two-mode squeezed states with low squeezing parameter,
permitting a loophole-free test of Bell inequalities.
\\
\end{abstract}
%We propose a Bell test for continuous variables based on homodyne
%detection using a conditionally prepared photon-number entangled
%state.  The state is near-optimal for Bell-inequality violations
%based on quadrature-phase homodyne measurements that use
%correlated photon-number states.  The scheme uses only beam
%splitters and photodetection to conditionally prepare a
%non-Gaussian state from a source of two-mode squeezed states with
%low squeezing parameter, permitting a loophole-free test of Bell
%inequalities.


\maketitle

Bell's theorem is regarded by some as one of the most profound
discoveries of science in the twentieth century.  Not only does it
provide a quantifiable measure of correlations stronger than any
allowed classically, which is a key resource in many quantum
information processing applications, it also addresses fundamental
questions in the foundations of quantum mechanics.  In 1964, Bell
quantified Bohm's version of the Einstein, Podolsky, and Rosen
(EPR) gedanken experiment, by introducing an inequality that
provides a test of local hidden variable (LHV) models
\cite{bell1964}.  A violation of Bell's inequality forces one to
conclude that, contrary to the view held by EPR, quantum mechanics
can not be both local and real. In order to experimentally support
this conclusion in a strict sense, a Bell test that is free from
loopholes is required.  Although it is still quite remarkable that
such seemingly metaphysical questions can even be put to the test
in the laboratory, a loophole-free Bell test has yet to be
achieved.

For more than three decades, numerous experiments have confirmed
the predictions of the quantum theory, thereby disproving local
realistic models as providing a correct description of physical
reality\cite{aspect1982}. However, all experiments performed to
date suffer from at least one of the two primary loopholes -- the
detection loophole and the locality loophole. The detection
loophole arises due to low detector efficiencies that may not
permit an adequate sampling of the ensemble space while the
locality loophole suggests that component parts of the
experimental apparatus that are not space-like separated could
influence each other.  The majority of Bell tests have used
optical systems to measure correlations, some achieving space-like
separations but still subjected to low efficiency photodetectors
(see, \textit{e.g.}, Ref. \cite{weihs1998}). Correlations in the
properties of entangled ions were shown to violate a Bell
inequality using high efficiency detectors eliminating the
detection loophole; however, the ions were not space-like
separated \cite{rowe2001}. A major challenge that has yet to be
achieved is to experimentally realize a single Bell test that
closes these loopholes.

The ease with which optical setups address the locality loophole
coupled with the currently achievable high efficiencies ($> 0.95$)
of homodyne detectors make Bell tests using quadrature-phase
measurements good candidates for a loophole-free experiment.
Furthermore, continuous quadrature amplitudes are the optical
analog of position and momentum and more closely resemble the
original state considered by EPR. Unlike photon counting
experiments which deal with the microscopic resolution of a small
number of photons, by mixing the signal with a strong field,
homodyne measurements allow one to detect a macroscopic current
\cite{reid1997}.

In this article, we propose a test of Bell inequalities using
homodyne detection on a conditional non-Gaussian ``source" state,
prepared using only passive optics and photon detection. Events
are pre-selected -- using event-ready detection one knows with
certainty that the desired source state has been produced --
requiring no post-processing. Photon detectors are only used in
the pre-selection process and only affect the probability of
successfully creating the source state whereas the actual
correlation measurements are performed using high efficiency
homodyne detectors.  The source is a correlated photon-number
state that is near-optimal for Bell tests using homodyne
detection, opening the possibility of a conclusive, loophole-free
test.

%%%%%%%%%%%%%%%%%%%%%%%%%%%%%%%%%%%%%%%%%%%%%%%%%%%%%%%%%%%%%%%%%
We consider a two-mode quantum state of light that can be written
as
\begin{equation}  \label{eq:correlated photon state}
    | \Psi \rangle = \sum_{n=0}^{\infty}  c_n  | n,n \rangle,
\end{equation}
which is correlated in photon number $| n,n \rangle = | n
\rangle_A \otimes | n \rangle_B$ for modes $A$ and $B$. For
example, the two-mode squeezed state $| \psi_\lambda \rangle$ has
coefficients given by $c_n=\lambda^n \sqrt{1-\lambda^2} $, where
$\lambda=\tanh(s)$ is determined by the squeezing parameter $s$
\cite{knight1985}. Such states are experimentally easy to
generate; however, because they possess a Gaussian Wigner
distribution in phase space, they are unsuitable for tests of Bell
inequalities using quadrature-phase measurements
%Bell showed that a nonnegative Wigner distribution acts as a local hidden variable model
%\cite{{bell1965}} for position and momentum.
as it is a requirement that the Wigner function possesses negative
regions \cite{{bell1964}}. Alternative, theoretically predicted
two-mode quantum superposition states called circle states, also
generated from vacuum fields through nondegenerate parametric
oscillation, having coefficients given by $c_n=r^{2n}/n!
\sqrt{I_0(2r^2)}$, do exhibit a violation for quadrature-phase
measurements with a maximum violation occurring for $r=1.12$
\cite{Gilchrist1998}. Unfortunately, unlike the two-mode squeezed
states, circle states are difficult to realize experimentally. A
recently proposed solution towards an experimentally realizable
state of light that is suitable for a homodyne Bell test is the
photon-subtracted two-mode squeezed state
\cite{Nha2004,GarciaPatron2004}, having coefficients
$c_n=\sqrt{(1-\lambda^2)^3/(1+\lambda^2)}(n+1)\lambda^n$, which
utilizes non-Gaussian operations on a Gaussian state.  In this
scheme, a photon is detected from each mode of a two-mode squeezed
state and only the resulting conditional state is used for
correlation measurements in the Bell test.  While the two-mode
squeezed state has a positive-everywhere Wigner function, the
conditional state after photon subtraction does not.

To date, all proposed states for a Bell test using
quadrature-phase measurements are not optimal states, meaning that
they do not produce the maximum possible violation of Bell
inequalities. The scheme presented here produces a two-mode
photon-entangled state that is near-optimal, using only beam
splitters and photon detection. The beam splitter may be described
by the unitary operator \cite{wodkiewicz1985}
\begin{equation}  \label{eq:BS unitary}
    U_{ab}=T^{a^\dag  a}e^{-R^* b^\dag a}e^{R b
    a^\dag}T^{-b^\dagger b},
\end{equation}
which describes the mixing of two modes $a$ and $b$ at a beam
splitter with transmissivity $T$ and reflectivity $R$. On-off
photon detection is described by the positive operator-valued
measure (POVM) of each detector, given by
\begin{equation}
    \Pi_0 = |0 \rangle \langle 0 |, \hspace{.2in} \Pi_1 = I-|0 \rangle \langle 0
    |.
\end{equation}
The on-off detectors distinguish between vacuum and the presence
of any number of photons.  The procedure is event-ready, a term
introduced by Bell, in the sense that one has a classical signal
indicating whether a measurable system has been produced.  The
states demonstrating a violation of local realism presented here
do not rely on the production of exotic states of light; in fact,
only a parametric source generating states with a low squeezing
parameter is required, making the procedure experimentally
feasible with current technology.  As depicted by Fig. 1, there
are three parties involved: Alice, Bob, and Sophie.  Sophie
prepares the source states that are sent to Alice and Bob, who
perform correlation measurements. We first describe the procedure
Sophie uses to generate the source states, which is shown by the
diagram in Fig. 2, and then discuss the measurements performed by
Alice and Bob.
%%%%%%%%%%%%%%%%%%%%%%%%%%%%%%%%%%%%%%%%%%%%%%%%%%%%%%%%%%%%%%%%%%%%%
    \begin{figure}[htbp]  \label{fig:f1BellSchematic8}
    \centerline{\scalebox{0.48}{\includegraphics{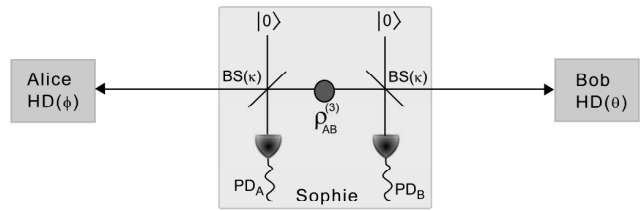}}}
    \caption[short caption.]
    {\small {A schematic diagram of the conditional homodyne detection
    for a Bell test.  Sophie records successful preparation of the source
    state, denoted by
    $|\psi^{(3)}\rangle_{PS},$ while Alice and Bob perform homodyne measurements at
    their space-like separated locations.  The three parties agree on which events
    to discard based on their measurement records.} }
    \end{figure}
%%%%%%%%%%%%%%%%%%%%%%%%%%%%%%%%%%%%%%%%%%%%%%%%%%%%%%%%%%%%%%%%%%%%%

In the first step, two-mode squeezed states are mixed pairwise at
unbalanced beam splitters followed by the non-Gaussian operation
associated with the POVM element $\Pi_1$. Specifically, a
non-Gaussian state is generated by
\begin{equation}  \label{eq:BellFockStateOperation}
    ( \Pi_{1,c} \otimes \Pi_{1,d})
    (U_{ac} \otimes U_{bd}) |\psi_\lambda \rangle
    |\psi_\lambda \rangle,
\end{equation}
where $|\psi_\lambda \rangle$ denotes the two-mode squeezed state
with $c_n=\lambda^n \sqrt{1-\lambda^2}$. For sufficiently small
$\lambda$, the operator $\Pi_1$ describing the presence of photons
at the detector approaches the rank-one projection onto the single
photon number subspace $|1\rangle\langle 1|$, which is still a
non-Gaussian operation. Under this condition, (un-normalized)
states of the form
\begin{equation}  \label{eq:BellFockState}
    | \psi^{(0)} \rangle = | 0,0 \rangle + \xi | 1,1 \rangle
\end{equation}
can be produced.  That is, even though the output state of
(\ref{eq:BellFockStateOperation}) will in general be a mixed
state, when $\lambda \in [0,1)$ is very small, the resulting
states can be made arbitrarily close in trace-norm to an entangled
state with state vector given by Eq. (\ref{eq:BellFockState}),
provided the appropriate choice of beam splitter transmittivity
$|T(\lambda)|=|\xi-\sqrt{\xi^2+8 \lambda^2}|/4 \lambda$ is used
\cite{browne2003}.  It should be emphasized that the state $|
\psi^{(0)}\rangle$, having a Bell-state form, can be generated for
arbitrary $\xi$.

It is interesting to note that the state given by Eq.
(\ref{eq:BellFockState}) does not violate a Bell inequality for
quadrature-phase measurements for any $\xi$, even when it has the
form of a maximally entangled Bell state, as was shown in Ref.
\cite{munro1999}, in which a numerical study of the optimal
coefficients for Eq. (\ref{eq:correlated photon state}) was
performed.  For certain values of $\xi$, Eq.
(\ref{eq:BellFockState}) describes a state that possesses a Wigner
distribution that has negative regions, showing that negativity of
the Wigner function is a necessary but not sufficient condition
for a violation of Bell inequalities using quadrature-phase
measurements.

The second step is to combine two copies of the state given by Eq.
(\ref{eq:BellFockState}) pairwise and locally at 50:50 beam
splitters described by the unitary operator of Eq. (\ref{eq:BS
unitary}). Detectors that distinguish only between the absence and
presence of photons are placed at the output port of each beam
splitter and when no photons are detected, the state is retained.
The resulting un-normalized state is
\begin{equation}
    |\psi^{(i+1)} \rangle = \langle 0,0 | U_{ac} \otimes U_{bd}
    | \psi^{(i)} \rangle | \psi^{(i)} \rangle = \sum_{n=0}^\infty
    c_n^{(i+1)}
    |n,n \rangle,
\end{equation}
where the coefficients are given by \cite{opatrny2000}
\begin{equation}
    c_n^{(i+1)}=2^{-n} \sum_{r=0}^n
    \left(
    \begin{array}{c}
    n \\
    r
    \end{array}
    \right)
    c_{r}^{(i)} c_{n-r}^{(i)}.
\end{equation}
It is optimal to iterate this procedure three times so that Sophie
prepares the state $|\psi^{(3)} \rangle$.  Each iteration leads to
a Gaussification of the initial state
\cite{browne2003,eisert2004}, which builds up correlated photon
number pairs in the sum of Eq. (\ref{eq:correlated photon state}).
Further iterations would Gaussify the state too much and destroy
the nonlocal features for phase space measurements.

The final step is to reduce the vacuum contribution by subtracting
a photon from each mode of the state $|\psi^{(3)} \rangle$,
obtaining a state proportional to $ab |\psi^{(3)} \rangle$. This
is done by mixing each mode with vacuum at two beam splitters with
low reflectivity.  A very low reflectivity results in single
photon counts at each detector with a high probability when a
detection event has occurred. Thus, the unitary operation
describing the action of the beam splitter is expanded to second
order in the reflectivity and the state is conditioned on the
result $N=1$ at each detector. The final photon-subtracted state,
given by
\begin{equation}
    |\psi^{(3)} \rangle_{PS}= {\mathcal N} \sum_{n=0}^\infty (n+1)c^{(3)}_{n+1}
    |n,n \rangle,
\end{equation}
where ${\mathcal N}$ is a normalization factor, is a near-optimal
state for homodyne detection.  Figure 3 compares the previously
proposed states -- the circle state and the photon-subtracted
two-mode squeezed state -- with the near-optimal state
$|\psi^{(3)} \rangle_{PS}$, as well as the numerically optimized
state in Ref. \cite{munro1999}.  The conditioning procedure alters
the photon number distribution of the input state and behaves
similarly to entanglement distillation.
%%%%%%%%%%%%%%%%%%%%%%%%%%%%%%%%%%%%%%%%%%%%%%%%%%%%%%%%%%%%%%%%%%%%%
    \begin{figure}[htbp]  \label{fig:f2BellTree}
    \centerline{\scalebox{0.53}{\includegraphics{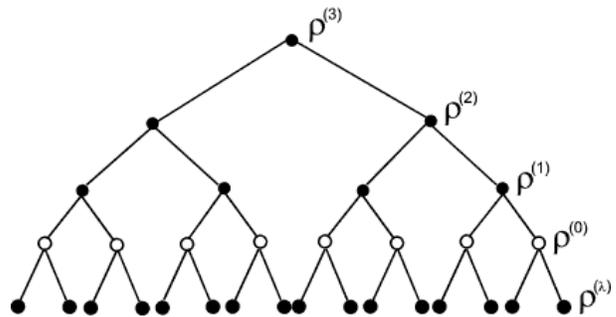}}}
    \caption[short caption.]
    {\small {A tree diagram of the conditional preparation of the state denoted by
    $|\psi^{(3)}\rangle$ from a finite supply of two-mode squeezed states $|\psi^{(\lambda)}\rangle$.
    Each circle represents
    the two-mode state as given on the right.  The black (white) circles
    correspond to the Gaussian (non-Gaussian) operations used to produce the states at
    each node.} }
    \end{figure}
%%%%%%%%%%%%%%%%%%%%%%%%%%%%%%%%%%%%%%%%%%%%%%%%%%%%%%%%%%%%%%%%%%%%%

Although the procedure used to create the correlated photon source
is probabilistic, with the success probability determined by the
amount of two-mode squeezing and the transmittivity of the
unbalanced beam splitters, it is event-ready -- Sophie has a
record of when the source state was successfully prepared. Low
efficiency photon detectors used in the state preparation only
affect the success probability and do not constitute a detection
loophole. Each mode of the source state $|\psi^{(3)}\rangle_{PS}$
is distributed to a separate location where correlation
measurements using high efficiency homodyne detectors are
performed by the two distant (space-like separated) parties, Alice
and Bob. Alice and Bob each mix their light modes with independent
local oscillators (LO) and randomly measure the relative phase
between the beam and the LO, taking into account the timing
constraint that ensures fair sampling. Alice measures the rotated
quadrature $x_{\theta}^A=x^A \cos \theta + p^A \sin \theta $ and
Bob measures the rotated quadrature $x_{\phi}^B=x^B \cos \phi +
p^B \sin \phi.$ Correlations are considered for two choices of
relative phase: $\theta_1$ or $\theta_2$ for Alice and $\phi_1$ or
$\phi_2$ for Bob. Finally, Alice, Bob and Sophie compare their
experimental results to determine when the source state was
successfully generated and which correlation measurements to use
for the Bell inequalities.

%%%%%%%%%%%%%%%%%%%%%%%%%%%%%%%%%%%%%%%%%%%%%%%%%%%%%%%%%%%%%%%%%%%%%
Two types of Bell inequalities will be examined -- the
Clauser-Horne-Shimony-Holt (CHSH) and Clauser-Horne (CH)
inequalities \cite{clauser1969}.  To apply these inequalities,
which are for dichotomous variables, the measurement outcomes for
Alice and Bob are discretized by assigning the value $+1$ if $x
\geq 0$ and $-1$ if $x<0$. Let $P^{AB}_{++}(\theta,\phi)$ denote
the joint probability that Alice and Bob realize the value $+1$
upon measuring $\theta$ and $\phi$, respectively and
$P^{A}_{+}(\theta)$ denote the probability that Alice realizes the
value $+1$ regardless of Bob's outcome, with similar notation for
the remaining possible outcomes. From LHV theories, the following
joint probability distribution can be derived:
\begin{equation}
    P^{AB}_{ij}(\theta,\phi)=\int \rho(\lambda) p_i^A(\theta,\lambda)
    p_j^B(\phi,\lambda) d \lambda
\end{equation}
with $i,j=\pm$, by postulating the existence of hidden variables
$\lambda$ and independence of outcomes for Alice and Bob. Quantum
mechanically, the joint probability distribution is given by the
Born rule $P(x^A_\theta,x^B_\phi)=|\langle x^A_\theta,x^B_\phi
|\psi^{(3)} \rangle_{PS}|^2.$  The probability for Alice and Bob
to both obtain the value $+1$ is
$P^{AB}_{++}(\theta,\phi)=\int_0^\infty \int_0^\infty
P(x^A_\theta,x^B_\phi) d x^A_\theta d x^B_\phi$. The joint
distribution is symmetric and a function of only the sum of the
angles $\chi=\theta+\phi$ permitting the identification
$P^{AB}_{++}(\theta,\phi)=P^{AB}_{++}(\chi)=P^{AB}_{++}(-\chi)$
and $P^{AB}_{++}(\chi)=P^{AB}_{--}(\chi)$. The marginal
distributions $P_+^{A}(\theta)=P_+^{B}(\phi)=1/2$ are independent
of the angle. Given the probability distributions, the predictions
of quantum theory can be tested with those of LHV theory.

First, we consider the Bell inequality of the CHSH type, which
arises from linear combination of correlation functions having the
form
\begin{equation}        \label{eq:bellcombo}
    \emph{B}= E(\theta_1,\phi_1)+E(\theta_1,\phi_2)+E(\theta_2,\phi_1)-E(\theta_2,\phi_2),
\end{equation}
where $E(\theta_i,\phi_j)$ is the correlation function for Alice
measuring $\theta_i$ and Bob measuring $\phi_j$. These
correlations are in turn determined by
\begin{equation}        \label{eq:correlation function}
    E(\theta,\phi)=P^{AB}_{++}(\theta,\phi)+P^{AB}_{--}(\theta,\phi)
    -P^{AB}_{+-}(\theta,\phi)-P^{AB}_{-+}(\theta,\phi),
\end{equation}
obtained through the many measurements that infer the
distributions $P^{AB}_{ij}(\theta,\phi)$.  With the aid of the
symmetry and angle factorization properties, the CHSH inequality
takes the simple form $\emph{B}= 3 E(\chi)-E(3 \chi)$ with LHV
models demanding that $|\emph{B}| \leq 2$.  The strongest
violation of the inequality is obtained for the value
$\chi=\pi/4$, thus, a good choice of relative phases for Alice and
Bob's measurements is $\theta_1=0$, $\theta_2=\pi/2,$
$\phi_1=-\pi/4$, and $\phi_2=\pi/4$.  Using homodyne detection
with optimal correlated photon number states, the maximum
achievable violation is 2.076 whereas using the source states
presented here, a Bell inequality violation of $\emph{B}=2.071$ is
achievable.

%%%%%%%%%%%%%%%%%%%%%%%%%%%%%%%%%%%%%%%%%%%%%%%%%%%%%%%%%%%%%%%%%%%%%
Let us also consider the Clauser-Horne (strong) Bell inequality
formed by the linear combination
\begin{equation}        \label{eq:chcombo}
    \frac{P^{AB}_{++}(\theta_1,\phi_1)-P^{AB}_{++}(\theta_1,\phi_2)+
    P^{AB}_{++}(\theta_2,\phi_1)+P^{AB}_{++}(\theta_2,\phi_2)}
    {P_+^{A}(\theta_2)+P_+^{B}(\phi_1)},
\end{equation}
denoted by $\emph{S}$, for which local realism imposes the bound
$|\emph{S}| \leq 1$. Again, using the properties of the
probability distributions, the simplification $\emph{S}= 3
P^{AB}_{++}(\chi)-P^{AB}_{++}(3 \chi)$ is possible.  With the
following choice of the phases: $\theta_1=0,$ $\theta_2=\pi/2$,
$\phi_1=-\pi/4$, and $\phi_2=\pi/4$, a violation of
$\emph{S}=1.018$ is attainable given the states in Eq.
(\ref{eq:BellFockState}) with parameter value $\xi=1/\sqrt{2}$,
which is quite close to the maximum value of 1.019 achieved by the
numerical, optimal states in Ref. \cite{munro1999} .
%%%%%%%%%%%%%%%%%%%%%%%%%%%%%%%%%%%%%%%%%%%%%%%%%%%%%%%%%%%%%%%%%%%%%
    \begin{figure}[htbp]  \label{fig:OptimalStateCompare}
    \centerline{\scalebox{0.8}{\includegraphics{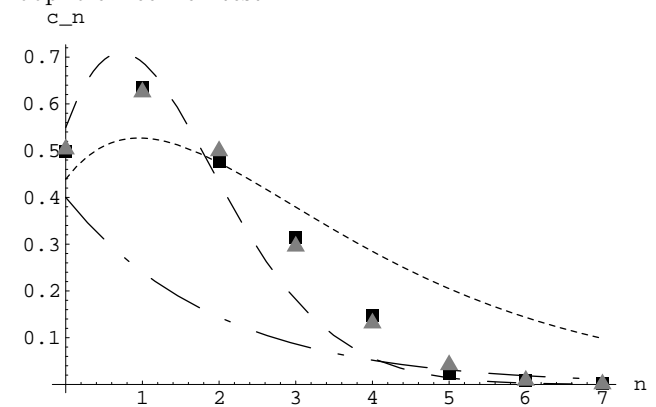}}}
    \caption[short caption.]
    {\small {The coefficients in Eq. (\ref{eq:correlated photon state}) are
    plotted as a function of photon number $n$.
    The curves represent photon-subtracted state two-mode squeezed state (small dashes)
    for $\lambda=0.6$,
    the circle states (large dashes) for $r=1.12$, and the two-mode squeezed state
    (dashed-dotted) for $\lambda=0.6$.
    The optimal states are shown by squares and the source states $|\Psi^{(3)}\rangle_{PS}$ are shown
    by grey triangles for $\xi=0.71$. } }
    \end{figure}
%%%%%%%%%%%%%%%%%%%%%%%%%%%%%%%%%%%%%%%%%%%%%%%%%%%%%%%%%%%%%%%%%%%%%

%%%%%%%%%%%%%%%%%%%%%%%%%%%%%%%%%%%%%%%%%%%%%%%%%%%%%%%%%%%%%%%%%%%%
We have shown how it is possible to prepare a near-optimal
state for a Bell test that uses quadrature-phase homodyne
measurements. Only very low squeezed states, passive optical
elements and photon detectors are required,  making the procedure
experimentally feasible at present.
%The state presented here is very close to the
%ideal Bell state for two-mode continuous variables in phase space,
%the ideal Bell state being the one that is most nonlocal.  As
%such, it may yield an improvement in teleportation schemes for
%continuous variables, for which the state used to execute the
%teleportation must be nonlocal (see, \textit{e.g.}, Ref.
%\cite{opatrny2000}).
An initial state with a positive-everywhere Wigner function was
succeeded by both non-Gaussian and Gaussian operations to prepare
a state that exhibits a strong violation of both the CHSH and CH
Bell inequalities.  Efforts are currently being made towards an
experimental realization of entanglement distillation for Gaussian
states.  The procedure presented here offers the opportunity for
another possible experiment, as it utilizes a subset of an
entanglement distillation procedure. Of course, any observed
violation of a Bell inequality is sensitive to inefficiencies in
the experiment that tend to deplete correlations.  A full analysis
involving dark counts and detection inefficiencies as addressed in
Ref. \cite{eisert2004} is necessary. Near-optimal states for
homodyne detection may allow a larger window for experimental
imperfections and offer the opportunity for a conclusive,
loophole-free Bell test.

%%%%%%%%%%%%%%%%%%%%%%%%%%%%%%%%%%%%%%%%%%%%%%%%%%%%%%%%%%%%%%%%%
We thank Bill Munro and Stefan Scheel for useful comments and
discussions. This work was supported by the U.S. National Science
Foundation under the program NSF01-154, by the U.K. Engineering
and Physical Sciences Research Council, and the European Union.

%%%%%%%%%%%%%%%%%%%%%%%%%%%%%%%%%%%%%%%%%%%%%%%%%%%%%%%%%%%%%%%%%%

\end{document}